\documentclass[12pt]{iopart}

\usepackage{epsf}
\def\ba{\begin{eqnarray}}
\def\ea{\end{eqnarray}}
\newcommand{\et}{{\it et al.~}}

\begin{document}

\title[Hamiltonian relaxation]{Hamiltonian relaxation}

\author{Pedro Marronetti}

\address{Department of Physics,
Florida Atlantic University, Boca Raton, FL 33431 \\
and \\
Department of Physics,
University of Illinois at Urbana-Champaign, Urbana, IL 61801}

\begin{abstract}
Due to the complexity of the required numerical codes, many of the new formulations for the evolution of the gravitational fields in numerical relativity are not tested on binary evolutions. We introduce in this paper a new testing ground for numerical methods based on the simulation of binary neutron stars. This numerical setup is used to develop a new technique, the Hamiltonian relaxation (HR), that is benchmarked against the currently most stable simulations based on the BSSN method. We show that, while the length of the HR run is somewhat shorter than the equivalent BSSN simulation, the HR technique improves the overall quality of the simulation, not only regarding the satisfaction
of the Hamiltonian constraint, but also the behavior of the total angular momentum of the binary. The latest quantity agrees well with post-Newtonian estimations for point-mass binaries in circular orbits.
\end{abstract}


\ead{pmarrone@physics.fau.edu}

\submitto{\CQG}

\maketitle


\section{Introduction}
\label{intro}

The numerical evolution of Einstein field equations has proved to be a formidable task. Ill-posed formulations of the dynamical equations, the presence of singularities, exponential growth of constraint violations, inadequate inner (in the case of black hole excision) and outer boundary conditions, and the lack of robust shock-handling hydrodynamical algorithms for strong gravity regimes are some of the open problems encountered in numerical relativity. To this list, we should add the problems intrinsic to the generation of astrophysically realistic initial data (ID) sets, such as the lack of adequate formulations to describe accurately the gravitational radiation content at the initial time step, the difficulty in specifying the fluid motion corresponding arbitrary stellar spins, or the inner boundary conditions around black hole singularities, to mention just a few. While none of the above issues has been solved to satisfaction, great progress has been made in the past decades in all these directions.

Controlling the exponential growth of constraint violating modes is essential for the stability of numerical simulations. Some of these modes are produced in the bulk of the numerical grid, originating in small violations due to roundoff and truncation errors which are inevitable in any numerical treatment. Other modes are generated at the grid boundaries by inappropriate boundary conditions (see, for instance, \cite{Lindblom:2004gd, Holst:2004wt}). While there is a purely numerical component to these instabilities, the scientific community consensus is that the choice of dynamical formulation (a given way of casting the Einstein field equations) will play an important role in their control \cite{Kidder:2001tz,Lehner:2001wq,Laguna:2002zc}. Most of the earliest attempts at modeling numerically general relativistic systems were based on a formulation introduced by Arnowitt, Deser, and Misner \cite{adm62}, widely known as ADM. However, it soon became clear that the life of the simulations could be extended by modifying ADM or simply replacing it with different formalisms. A method originally developed by Shibata and Nakamura \cite{Shibata:1995} and later used by Baumgarte and Shapiro \cite{Baumgarte:1998te} (BSSN) is today the most commonly used for three-dimensional simulations. The literature offers an ever-growing list of new evolution formulations which can be divided in two groups: unconstrained \cite{Kidder:2001tz,Laguna:2002zc,Shibata:1995,Baumgarte:1998te,Bondi:1962aa,Sachs:1962aa,bona98,Anderson:1999qx,Arbona:1999ym,Bishop:1999yg,Gentle:2003hs,Bona:2003qn,Tiglio:2003cf,Tiglio:2003xm,Bishop:2003bs} and constrained \cite{Bonazzola:2003dm,Anderson:2003dz}. Some of these have been tested numerically under conditions that are either easy to implement numerically and / or have a high degree of symmetry. Test cases have been constructed based on perturbations of Minkowski spacetime, linearized waves, non-linear plane waves, Gowdy and Robertson-Walker metrics, Brill and Klein-Gordon waves, non-linear versions of Maxwell equations, and single black hole spacetimes (for a comprehensive analysis of such tests, we refer the reader to \cite{Alcubierre:2003pc}). The importance of these tests is twofold: they show the weaknesses and strengths of each formulation in easy to understand cases and they serve as a first discriminatory round of benchmarks that are simple enough to be in the reach of most research groups. However, binary simulations have been performed sparsely: they require very complex numerical codes as well as non-trivial computational power. To our knowledge, ADM and BSSN are the only formulations that have been tested in three-dimensional finite-size compact-object binary simulations: either with neutron stars \cite{Shibata:1999wm,Shibata:2002jb,Duez:2002bn,Shibata:2003ga,Miller:2003vc,Marronetti:2003hx,Kawamura:2003hu} or black holes \cite{Brugmann:1997uc,Brandt:2000yp,Alcubierre:2000ke,Bruegmann:2003aw,Alcubierre:2004hr}.

In this paper, we design and implement a numerical setup based on binary neutron stars (BNS) simulations. BNS inspiral scenarios are fully general relativistic, three-dimensional, and do not present singularities, thus decoupling the problem of inner boundary conditions needed for black hole excision. They provide a robust test for the capacity of the numerical implementation to handle angular momentum conservation, usually one of the most problematic quality control monitors \footnote[1]{Total angular momentum is not strictly conserved in general relativity. However, most of the variations observed in current simulations are due to numerical error and not to the physical loss due to gravitational radiation.}. They also are astrophysically realistic systems which are related to some of the most important observational phenomena of our times: gamma-ray burst engines and generators of gravitational waves. BNS simulations usually require extensive computational resources and the length of the runs could, in principle, render these tests impractical. Here we show how small, low resolution grids can be used to gain insight into the stability of different numerical schemes, with runs that only take a few hours on single-processor workstations. 

To illustrate the efficacy of the proposed testing ground, we introduce a new technique that was developed with the help of short trial-and-error BNS simulations. This method is based on the approximate satisfaction of the Hamiltonian constraint, obtained by relaxing the conformal factor. Anderson and Matzner \cite{Anderson:2003dz} recently performed black-hole simulations where the Hamiltonian and momentum constraints elliptic equations are solved at every time step. In our approach (which for this paper involves only the Hamiltonian constraint) the conformal factor is relaxed from a parabolic equation that drives the Hamiltonian constraint into exponentially decaying solutions. This method is an adaptation of the K-Driver algorithm introduced by Balakrishna \et \cite{betal96}, used to enforce the maximal slicing condition. This technique reduces the Hamiltonian constraint violation by a factor of five with respect to the ID values, and by two orders of magnitude with respect to BSSN simulations. Berger \cite{Berger:2004dd} has shown recently how the satisfaction of the Hamiltonian constraint throughout the evolution plays a role in the overall quality of the simulation. We observe a similar effect in our runs which is particularly evident in the remarkable improvement of the behavior of the total angular momentum of the binary. The total angular momentum as a function of time agrees with post-Newtonian estimations for more than one orbital period for the case of BNS outside the innermost stable circular orbit (ISCO). In addition, the computational overhead caused by the relaxation method adds only an extra 5\% to the running time of the simulations, making the technique more practical than solving the elliptic equation derived from the Hamiltonian constraint.

Section \ref{EE} describes the time evolution equations for the gravitational fields, summarizing the differences between BSSN and the Hamiltonian relaxation method with special emphasis in the boundary conditions employed in this paper. Section \ref{NB} describes the BNS numerical experiment, introducing the new full general relativistic hydrodynamical code (GRHyd) employed for the time evolution. Section \ref{results} presents results obtained with the Hamiltonian relaxation method and compares them with the corresponding BSSN counterparts. The Appendix section shows validation and convergence tests, as well as the derivation of the post-Newtonian (PN) calculation for the loss of angular momentum for point-mass binaries.


\section{Equations for the Gravitational and Hydrodynamical Fields}
\label{EE}

\subsection{Time Evolution Equations}
\label{TEE}

In this paper we use geometrized units ($G=c=1$) and the Greek (Latin) indices run from 0 to 3 (1 to 3). 

Numerical relativistic treatments usually cast the metric in the ``3+1" form
\ba
\rmd s^2 = -\alpha^2 \rmd t^2 + \gamma_{ij}(\rmd x^i+\beta^i \rmd~t)(\rmd x^j+\beta^j \rmd t)~,
\ea
where $\alpha$, $\beta^i$, and $\gamma_{ij}$ are the lapse function, shift vector, and spatial metric tensor, respectively. The extrinsic curvature $K_{ij}$ is defined as
\ba
K_{ij} = -(\partial_t - {\mathcal{L}}_{\beta})\gamma_{ij}~/ ~(2 \alpha)~,
\ea
where ${\mathcal{L}}_{\beta}$ is the Lie derivative with respect to $\beta^i$ \cite{Baumgarte:1998te}. Using these fields, we can rewrite Einstein's field equations
\ba
G_{\mu\nu} = 8 \pi T_{\mu\nu}
\ea
as a set of four differential equations:
\ba
R - K_{ij}K^{ij} + K^2 = 16\pi\rho~, \\
\label{HC}
D_jK^j_i - D_iK = 8\pi S_i~,
\label{MC}
\ea
known as the Hamiltonian and momentum constraints, and
\ba
\fl (\partial_t - {\mathcal{L}}_{\beta})\gamma_{ij} = -2 \alpha K_{ij}~,\nonumber \\
\fl (\partial_t - {\mathcal{L}}_{\beta}) K_{ij}=
        - D_iD_j\alpha + \alpha \{ R_{ij}  -  2K_{il}K^l{}_j + K K_{ij}
        - 8\pi [S_{ij}  +  \frac{1}{ 2} \gamma_{ij} (\rho-S)]\}~,
\label{ADM}
\ea
which provide the evolution in time of the spatial metric and the extrinsic curvature. The symbol $D_i$ represents the covariant gradient with respect to the tensor $\gamma_{ij}$. The fields $\rho$, $S$, and $S_{ij}$ are derived from the matter fields by splitting the stress-energy tensor $T_{\mu\nu}$ in components parallel and perpendicular to the normal of the spatial hypersurface $n^\alpha$
\ba
\rho = n_{\alpha}n_{\beta}T^{\alpha\beta}~, \nonumber \\
S_i  = -\gamma_{i\alpha}n_{\beta}T^{\alpha\beta}~, \nonumber \\
S_{ij} = \gamma_{i\alpha}\gamma_{j\beta}T^{\alpha\beta}~, \nonumber \\
S  = \gamma^{ij} S_{ij}~. \nonumber
\ea
These equations are the basis of the ADM formulation \cite{adm62}. 

Following York \cite{y79}, we rewrite the metric and the extrinsic curvature as
\ba
\gamma_{ij} = \psi^4 ~\tilde{\gamma}_{ij}~, \nonumber\\
K_{ij} = \psi^4 ~\bigl(\tilde A_{ij} + \frac{1}{3} \tilde\gamma_{ij}K\bigr)~. \nonumber
\ea
These decompositions define the fields $\psi$, $\tilde{\gamma}_{ij}$, $\tilde A_{ij}$, and $K$, known as the conformal factor, the conformal metric, the conformal traceless extrinsic curvature, and the trace of the extrinsic curvature respectively. Using these variables, the Hamiltonian and momentum constraints can be rewritten as
\ba
\label{HC2}
  \tilde\gamma^{ij}\tilde D_i\tilde D_j \psi 
           - \frac{\psi}{8}\tilde R 
	   + \frac{\psi^5}{8}\tilde A_{ij}\tilde A^{ij}
           -  \frac{\psi^5}{12}K^2 
	   +  2\pi \psi^5 \rho  =  0, \\
\label{MC2}
  \tilde D_j(\psi^6 \tilde A^{ji})- \frac{2}{3} \psi^6 \tilde D^i K
  - 8\pi \psi^6 S^i
  = 0~.
\ea
The ID sets are generated by solving (usually numerically) these equations. These solutions are correct within the bounds of the truncation error associated with the finite-difference scheme of choice. These initial constraint violations will be amplified when using unconstrained formulations like ADM and BSSN.

Equations for the time evolution of the fields $\tilde{\gamma}_{ij}$, $K$, and $\tilde A_{ij}$ can be derived from (\ref{ADM}):
\ba
\fl (\partial_t - {\mathcal{L}}_{\beta})\tilde\gamma_{ij}
                = -2\alpha\tilde A_{ij} \nonumber \\
\fl (\partial_t - {\mathcal{L}}_{\beta})K
                = -\gamma^{ij}D_jD_i\alpha + \frac{1}{3}\alpha K^2 
                  + \alpha \tilde A_{ij}\tilde A^{ij}
                    + 4\pi\alpha (\rho + S) \nonumber \\
\fl (\partial_t - {\mathcal{L}}_{\beta})\tilde A_{ij}
                = \psi^{-4} [-D_iD_j\alpha
                    + \alpha(R_{ij}-8\pi S_{ij})]^{TF}
                  + \alpha(K\tilde A_{ij} - 2\tilde A_{il}\tilde A^l{}_j)~,
\label{dot_eqs}
\ea
where the superscript $TF$ indicates the trace-free part of the tensor. These fields are complemented with the variable known as the conformal connection, introduced in \cite{Shibata:1995}
\ba
\tilde\Gamma^i  \equiv  -\tilde\gamma^{ij}{}_{,j},
\ea
where we follow the notation of \cite{Baumgarte:1998te}. An evolution equation is derived for these variables from (\ref{MC2}) and (\ref{dot_eqs}):
\ba
\fl \partial_t\tilde\Gamma^i &= \partial_j(2\alpha\tilde A^{ij} 
                + {\mathcal{L}}_{\beta}\tilde\gamma^{ij}) \nonumber \\
\fl      &=  \tilde\gamma^{jk}\beta^i{}_{,jk} 
            + \frac{1}{3}\tilde\gamma^{ij}\beta^k{}_{,kj}
            - \tilde\Gamma^j\beta^i{}_{,j}
	    + \frac{2}{3}\tilde\Gamma^i\beta^j{}_{,j} 
            + \beta^j\tilde\Gamma^i{}_{,j} 
	    - 2\tilde A^{ij}\partial_j\alpha \nonumber \\
\fl       & - 2\alpha ~(\frac{2}{3} \tilde\gamma^{ij}K_{,j} 
            - 6\tilde A^{ij} ~[\ln(\psi)]_{,j} 
            - \tilde\Gamma^i{}_{jk}\tilde A^{jk} 
	    + 8\pi\tilde\gamma^{ij}S_j)~.
\label{Gamma_dot}
\ea
Equations (\ref{dot_eqs}) and (\ref{Gamma_dot}) together with a time evolution equation for the conformal factor
\ba
(\partial_t - {\mathcal{L}}_{\beta}) \ln(\psi) = - \frac{1}{6}\alpha K
\label{cf_dot_BSSN}
\ea
are the basis of the BSSN formulation \cite{Shibata:1995, Baumgarte:1998te}.

We define the Hamiltonian constraint residual $\mathcal{H}$ as the r.h.s. of equation (\ref{HC2}). Note that $\mathcal{H}$ will only be null when dealing with exact solutions in the continuum case; in any numerical treatment the round-off and truncation errors will violate the constraint (\ref{HC2}). This equation is an elliptic second-order PDE that, in principle, can be used to determine the conformal factor. Anderson and Matzner \cite{Anderson:2003dz} performed simulations of static single black-hole spacetimes, solving equation (\ref{HC2}) at every time step. 

The Hamiltonian relaxation technique proposed here derives the conformal factor from solving approximately the Hamiltonian constraint, in a way that resembles the K-Driver algorithm \cite{betal96} used for calculating the lapse function. Instead of solving equation (\ref{HC2}), we compose an alternative parabolic equation which will relax $\psi$ to a solution of the Hamiltonian constraint through an iterative scheme. An example of such equation is
\ba
\partial_{t'} \psi = \epsilon_H ~(\partial_{t'} \mathcal{H} + \eta_H ~\mathcal{H})~,
\label{psi_dot}
\ea
where $\epsilon_H$ and $\eta_H$ are fine-tuning parameters. Note that the time coordinate $t'$ is not the physical time, but a relaxation parameter that has only numerical meaning: for every $\Delta t$ of physical time, equation (\ref{psi_dot}) will be relaxed up to a maximum number of times $\Delta t / \Delta t'$ (see below). A stationary solution (in $t'$) of equation (\ref{psi_dot}) enforces the condition $\partial_{t'} \mathcal{H} = -\eta_H ~\mathcal{H}$, which corresponds to an exponential decay of the Hamiltonian constraint \cite{betal96}.

The $\partial_{t'} \mathcal{H}$ term of equation (\ref{psi_dot}) could in principle be derived from the time evolution equations (\ref{dot_eqs}). However, the resulting equation is quite long and difficult to handle numerically. A much simpler alternative is the use of the Forward-Time Centered-Space (FTCS) finite-difference expression
\ba
\partial_{t'} \mathcal{H} \simeq \frac{\Delta \mathcal{H}}{\Delta t'} = \frac{\mathcal{H}^{~m} - \mathcal{H}^{~m-1}}{\Delta t'}~,
\ea
where $\mathcal{H}^{~m}$ represents the constraint violation at the relaxation iteration step $m$.

The relaxation of $\psi$ is performed at each one of the steps of the ICN method (one Predictor and two Corrector steps) and is done after the corresponding update of the rest of the gravitational fields and before the update of the hydrodynamical fields (i.e.; $\rho$). The relaxation in the Predictor stage follows these steps:\\

1) Initial update of $\psi$:
\ba
\psi^0_n = \psi_{n-1} + \Delta t'~ \epsilon_H ~\eta_H ~ \mathcal{H}_{n-1},
\ea
where the subscript $n$ follows the physical time step. Field values from the previous time step (i.e.; $\psi_{n-1}$, $\mathcal{H}_{n-1}$, etc.) do not carry any upper index since they are not ``relaxed". This update is followed by a re-evaluation of the boundary conditions for $\psi^0_n$ which are explained in detail in section \ref{BCs}.\\

2) Subsequent updates of $\psi$: Loop on index $m$ from 1 to $M$
\ba
\psi^m_n = \psi^{m-1}_n + \Delta t'~ \epsilon_H \left( \frac{\mathcal{H}^{~m-1}_n - \mathcal{H}^{~m-2}_n}{\Delta t'}+~\eta_H ~ \mathcal{H}^{~m-1}_n \right),
\label{update}
\ea
where the Hamiltonian residual $\mathcal{H}^{~m-1}_n$ is evaluated as
\ba
\mathcal{H}^{~m-1}_n = \mathcal{H}^{~m-1}_n (\psi^{m-1}_n, \tilde{\gamma}_{ij~(n-1)}, \tilde{A}_{ij~(n-1)}, etc.)~.
\ea
These updates of $\psi^m_n$ are also followed by a re-evaluation of the boundary conditions, after which the $L_2$ norm $||\mathcal{H}^{~m}_n||_2$ is calculated . The relaxation iteration is stopped when $||\mathcal{H}^{~m}_n||_2 < ||\mathcal{H}_{n-1}||_2$ or when the maximum number of iterations steps $M$ (chosen here to be 25) is reached.

The same steps are followed in the first Corrector (second Corrector) stage, but replacing the field values from the previous time step with the corresponding updates generated in the Predictor (first Corrector) step.


\subsection{Lapse and Shift Equations}

The lapse function $\alpha$ and shift vector $\beta^i$ (i.e.; the gauge fields) are related to the coordinate degrees of freedom of the theory of general relativity. The choice of lapse function controls the way the spacetime continuum is split into a foliation of spatial hypersurfaces, while the shift vector indicates how the spatial coordinates change from one hypersurface to the next. The gauge fields are usually chosen to gain stability during the simulation, thus, they depend on the choice of time evolution methods and the particular physical system under study. Previous BNS studies \cite{Duez:2002bn,Marronetti:2003hx} have shown that the maximal slicing condition (namely, $K=0$) implemented with the K-Driver algorithm \cite{betal96} is a robust choice for the lapse function when using the BSSN formulation. Note that for our simulations maximal slicing conditions of any form have a clear advantage over other alternatives, given that the ID sets are based on this condition for the lapse. This condition is based on the parabolic equation for the lapse function $\alpha$
\ba
\partial_{t'} \alpha = -\epsilon_\alpha ~(\partial_{t'} K + \eta_\alpha K)~,
\label{K_driver}
\ea
where $\epsilon_\alpha$ and $\eta_\alpha$ are free parameters.

The same studies \cite{Duez:2002bn} also showed that the $\Gamma$-Driver \cite{ab01} condition for the shift vector makes possible long term BNS evolutions. A similar parabolic relaxation method is used here, this time to relax the shift vector $\beta^i$ in such a way that it drives the BSSN variable $\tilde{\Gamma}^i$ exponentially to zero. The corresponding parabolic equation is 
\ba
\partial_{t'} \beta^i = \epsilon_\beta ~ (\partial_{t'} \tilde\Gamma^i + \eta_\beta \tilde\Gamma^i)~,
\label{Gamma_driver}
\ea
where $\epsilon_\beta$ and $\eta_\beta$ are the corresponding free parameters. We also used in our simulations the condition known as $\beta$ freezing, that leaves the shift vector unchanged (i.e., identical to its initial value) all throughout the evolution. 

Finally, all the simulations presented in this paper were performed in the frame that rotates with the binary (corotating), which has been proved to enhance the stability of the evolutions not only in generally relativistic  \cite{Duez:2002bn, Marronetti:2003hx}, but also in Newtonian binary simulations \cite{Swesty:1999ke}.


\subsection{Boundary Conditions}
\label{BCs}

The choice of outer boundary conditions can be crucial for the stability of any numerical scheme. We adopt Sommerfeld (radiative) boundary conditions for the conformal metric $\tilde{\gamma}_{ij}$, the traceless part of the extrinsic curvature $\tilde{A}_{ij}$, and the trace of the extrinsic curvature $K$, and we set $\tilde{\Gamma}^i=0$ at the boundaries (see Duez \et \cite{Duez:2002bn} for details). These conditions are used for both BSSN and Hamiltonian Relaxation runs. 

\subsubsection{Boundary Conditions for the Conformal Factor $\psi$}

One important difference between the BSSN and the Hamiltonian relaxation runs is the boundary condition adopted for the conformal factor $\psi$. 
The BSSN runs use the same (Sommerfeld) boundary condition employed for the other gravitational fields. 

The Hamiltonian relaxation runs use a boundary condition that enforces the satisfaction of $\mathcal{H}=0$ in its finite-difference form. Equation (\ref{HC2}) can be expressed as
\ba
\tilde\gamma^{ij}\tilde D_i\tilde D_j \psi = 
\tilde\gamma^{ij}~\partial_i \partial_j \psi - \tilde\Gamma^k \partial_k \psi = \rho_\psi~,
\label{BC1}
\ea
where $\rho_\psi$ collects all the terms that do not depend on derivatives of $\psi$. Let's consider first the case of the grid points in the middle of the cube faces (i.e.; excluding the edges of the grid). If our numerical grid is a Cartesian cube with $N^3$ points, the points corresponding to the surface $x=x_{max}$ are represented by the indices $(N,j,k)$ \footnote[1]{In the remainder of this section, the indices $i$, $j$, and $k$ will represent inner grid points (i.e., $1<i,j,k<N$).}. As an example of our boundary condition for the conformal factor, we will derive an expression for the boundary values $\psi(x_{max})=\psi_{N,j,k}$, based on the finite-difference approximation of equation (\ref{BC1}) at the point next to the boundary (i.e.; the point with indices $(N-1,j,k)$). There, the differential operators involving partial derivatives along the $x$ direction are expressed as
\ba
\partial_x \psi^n & \sim & \frac{\psi_{N,j,k}^n-\psi_{N-2,j,k}^n}{2\Delta x}~, \nonumber \\
\partial_x \partial_x \psi^n & \sim & \frac{\psi_{N,j,k}^n-2 \psi_{N-1,j,k}^n+\psi_{N-2,j,k}^n}{(\Delta x)^2}~, \nonumber\\
\partial_x \partial_y \psi^n & \sim & \frac{\psi_{N,j+1,k}^n- \psi_{N-2,j+1,k}^n-\psi_{N,j-1,k}^n+\psi_{N-2,j-1,k}^n}{4~\Delta x \Delta y}~, \nonumber\\
\rm{etc,}\nonumber
\ea
where $n$ is the time step index and $\Delta x$ and $\Delta y$ represent the grid spacing along the $x$ and $y$ axis, respectively. Replacing the differential operators of equation (\ref{BC1}) with their finite-differencing counterparts results in an algebraic equation from which we obtain an expression for $\psi_{N,j,k}^n$ of the form
\ba
\psi_{N,j,k}^n = \frac{F_{N,j,k}(\psi^n) ~ (\Delta x)^2} {(\tilde{\gamma}_{xx})_{N-1,j,k}- \tilde{\Gamma}^x_{N-1,j,k}~\Delta x}~,
\label{psi_BC}
\ea
where $F_{N,j,k}(\psi^n)$ represents an algebraic function of $\psi_{i,j,k}^n$, $\psi_{N,j-1,k}^n$, $\psi_{N,j+1,k}^n$, $\psi_{N,j,k-1}^n$, and $\psi_{N,j,k+1}^n$. Equation {\ref{psi_BC}) defines a system of linear equations on the unknowns $\psi_{N,j,k}^n$, whose solution can be very time consuming. A faster alternative is to evaluate the equation (\ref{psi_BC}) using the values $\psi_{i,j,k}^n$, $\psi_{N,j-1,k}^n$, $\psi_{N,j+1,k}^{n-1}$, $\psi_{N,j,k-1}^n$, and $\psi_{N,j,k+1}^{n-1}$: we use the newly calculated values at the boundaries as soon as they are ready and use the previous step counterparts for the rest. In practice, $\psi^{n-1}$ corresponds to the last update of such boundaries values. Given the nature of the iterative Crank-Nicholson method, these updates occur three times in each time step, making our approach more accurate than expected.

This boundary condition for $\psi$ is essential in the implementation of the Hamiltonian relaxation method: tests performed using Sommerfeld conditions have shown a degradation of the quality of the simulation to BSSN levels. The same tests also show that the constraint violating noise generated at the boundaries can be reduced even further by generating the expression (\ref{psi_BC}) from the equation $\mathcal{H}_{N-1,j,k} = \mathcal{H}_{N-2,j,k}$, instead of $\mathcal{H}_{N-1,j,k} = 0$, which guarantees a smoother transition of the Hamiltonian residual at the next-to-the-boundary points $(N-1,j,k)$. 

A problem arises at the edges of the grid, where the expression for the boundary value $\psi_{N,N,k}^n$
\ba
\psi_{N,N,k}^n = \frac{F_{N,N,k}(\psi^n) ~ \Delta x ~ \Delta y} {2 ~ (\tilde{\gamma}_{xy})_{N-1,N-1,k}}~,
\label{psi_BC_edge}
\ea
is not well defined at the initial time step, when $\tilde{\gamma}_{xy}=0$. We experimented with several different alternatives and found that using a simple bilinear extrapolation of the form
\ba
\psi_{N,N,k} = \psi_{N-1,N,k} + \psi_{N,N-1,k} - \psi_{i,j,k}~,
\ea
when $\tilde{\gamma}_{xy}$ falls below some threshold, performs well for more than an orbital period.

\subsubsection{Boundary Conditions for the Gauge Fields $\alpha$ and $\beta^i$}

The BSSN runs were performed using Robin-like boundary conditions for the gauge fields, to be consistent with previous work \cite{Duez:2002bn,Marronetti:2003hx}. These conditions are based on the asymptotic behavior of the gauge fields at distances far from the matter sources \cite{Baumgarte:1997eg,Duez:2002bn}:
\ba
\alpha-1 &\propto& r^{-1}~, \nonumber \\
\beta^x  &\propto& y ~r^{-3} - y ~\omega ~, \nonumber \\
\beta^y  &\propto& x ~r^{-3} + x ~\omega ~, \nonumber \\
\beta^z  &\propto& x ~y ~z ~r^{-7} , \nonumber
\ea
where the terms proportional to the orbital angular velocity $\omega$ arise from working in the corotating frame ($\vec{\omega}=\omega ~\hat{z}$).

For the Hamiltonian relaxation runs we found that freezing the lapse function at the boundaries to its initial value results in a clear reduction of the incoming constraint violating noise. Hamiltonian relaxation test runs were performed using both the $\beta$-freeze condition and the $\Gamma$-Driver algorithm with Robin-like boundary conditions.


\subsubsection{Hydrodynamical Formulation}
The evolution of the gravitational fields is coupled with the corresponding hydrodynamical equations for the matter sources. For this reason, BNS testing is also a valid numerical setup for hydrodynamical formulations and their corresponding numerical implementations. In the same way the inspiral phase of the BNS offers ideal conditions for the testing of gravitational evolution algorithms, the merger phase can be used to discriminate between different hydrodynamical methods, with special interest in their ability to conserve angular momentum and capture shocks. For this paper, we adopt the hydrodynamical formulation and numerical techniques that have been used successfully in the simulation of the inspiral phase of BNS \cite{Duez:2002bn,Marronetti:2003hx}, since we are only interested in testing gravitational evolution schemes. We describe the fluid inside the stars using a perfect fluid stress-energy tensor and a polytropic equation of state, with constant $\Gamma=2$. Section \ref{VL} provides some characteristics of the hydrodynamical part of our code. For more details, we refer the reader to the above mentioned papers.


\section{Numerical Testing Ground}
\label{NB}

We design a testing ground for numerical algorithms that captures some of the main characteristics of compact-object binary scenarios: three-dimensionality, full general relativistic gravitation, and hydrodynamical evolution. Two types of numerical codes are necessary for this endeavor. One to generate astrophysically realistic BNS ID sets (CFC-Solver) and another to evolve this data in time (GRHyd). Our codes work on Cartesian grids and use finite-difference second order operators within grids that have uniform and identical spacing along each axis. The BNS considered in this paper are composed of identical stars with equatorial symmetry, to allow for the implementation of $\pi$ and equatorial symmetry in our codes. We work in the reference frame that rotates with the binary, and the stars are aligned with the $y$ axis.

\subsection{Initial Data Code: CFC-Solver}
\label{GRHyd}

The core of CFC-Solver is an elliptic solver based on a variation of a multigrid algorithm used for previous work  \cite{Marronetti:1998xv, Marronetti:1999ya}. It generates ID sets for BNS through the solution of the Hamiltonian and momentum constraints for a quasi-equilibrium circular orbit, via the Wilson-Mathews Conformally Flat Condition (CFC) approach \cite{Wilson:1995ty, Wilson:1996ty}. The CFC method simplifies the constraint equations by restricting the spatial metric tensor to a conformally flat form and favors ``orbit circularity" by imposing a helical Killing vector to the spacetime. CFC-Solver can generate data for stars with arbitrary masses and spins. For simplicity, the testing ground explored in this article includes only corotating and irrotational BNS. We refer the reader to \cite{Marronetti:2003gk} for more details about the code.

For this article, we use two different initial data sets: one describes a corotating BNS in a low-resolution small grid and the other an irrotational BNS in a high-resolution large grid. The latter is the same ID set used in previous work \cite{Marronetti:2003hx}, to facilitate the comparison of results obtained with our new code. Details of the ID sets are given in table \ref{table_grids}.

\subsection{Evolution Code: GRHyd}
\label{GRHyd}

The results presented in this paper were obtained with a new numerical code for the time evolution of gravitational and hydrodynamical fields. GRHyd was written using the Cactus programming environment \cite{Cactus}. One of the defining characteristics of the code is the separation of the formulations in different and interchangeable modules (``thorns" in Cactus language). Each module implements a given set of evolution equations, finite-difference schemes, gauge fields, and the corresponding boundary conditions. This applies to both the gravitational and hydrodynamical formulations. Comparisons between different gravitational formulations are performed employing the same hydrodynamical scheme. In this paper we will compare the results of the Hamiltonian relaxation scheme with those of BSSN, using in both cases a hydrodynamical module based on a van Leer advection scheme \cite{vl77} with artificial viscosity. 

\subsubsection{BSSN and Hamiltonian Relaxation Modules}
\label{BSSN}

The BSSN module, which follows closely the numerical techniques and parameters implemented by Duez \et \cite{Duez:2002bn} and Marronetti \et \cite{Marronetti:2003hx} for BNS simulations, is based on a second order in time iterative Crank-Nicholson (ICN) scheme. This module plays two essential roles: code validation and benchmarking of new formulations. To validate the new code GRHyd, we use the BSSN and van Leer modules to perform a simulation based on one of the ID sets used in \cite{Marronetti:2003hx} (see \ref{appendix_CT} for the comparison of results). 

Any new numerical method for the evolution of gravitational fields will 
be matched against BSSN simulations. Thus, the BSSN module provides the reference frame on which the new schemes will be measured. The first level of comparison involves short trial-and-error runs, where poorly performing schemes can be quickly weeded out. After this first round, the best performing methods will be used in high-resolution large-grid runs, to test their effectiveness in more realistic and demanding simulations.

The Hamiltonian relaxation module implements the algorithm described in section \ref{EE}. The values of the relaxation parameters used in this paper $\epsilon_H=0.0001$ and $\eta_H = 70.0$, were determined empirically. The relaxation is performed until the $L_2$ norm of the Hamiltonian residual is smaller than the one at the previous time step for up to a maximum of 25 iterations.

Both BSSN and Hamiltonian relaxation modules use ICN with a Courant factor of $0.46$. They use the same boundary conditions for the fields $\tilde{\gamma}_{ij}$, $\tilde{A}_{ij}$, $K$ (Sommerfeld), and $\tilde{\Gamma}^i$ ($\tilde{\Gamma}^i=0$). They also share the same K-Driver and $\Gamma$-Driver parameters: $\epsilon_\alpha=0.125$ 
\footnote[1]{The value $\epsilon_\alpha=0.625$ reported in \cite{Duez:2002bn} is incorrect.}, $\eta_\alpha=0.1$, $\epsilon_\beta=0.0005$, $\eta_\beta=0.2$. The K-Driver ($\Gamma$-Driver) relaxation is iterated 5 (10) times. 
The modules differ, however, in the boundary conditions for the lapse: the BSSN module employs Robin-like conditions, while the Hamiltonian relaxation freezes the lapse at the boundaries to its initial value.\\

\label{HR}

\subsubsection{Van Leer Module}
\label{VL}

The van Leer module implements the evolution of the hydrodynamical fields using an ICN scheme that couples to the evolution of the gravitational fields. The fluid advection follows a van Leer algorithm \cite{vl77} that is complemented with the introduction of artificial viscosity, to handle the presence of shocks. This method, while not as  sophisticated as high-resolution shock-handling techniques \cite{HRSC}, is fast, easy to program, and good enough for the simulation of the BNS inspiral phase as long as no shocks are present. During the inspiral, shocks can develop only in the atmospheric envelope that surrounds the stars, and the numerical handling of such atmosphere usually requires a good deal of work \cite{Swesty:1999ke}. We avoid shocks and the problems associated with them by adopting the non-atmospheric method introduced in \cite{Duez:2002bn}. We use {\it Copy} \cite{Duez:2002bn} outer boundary conditions for the hydrodynamical fields. 

\subsection{Benchmark Runs}
\label{BR}

For the short trial-and-error runs to be practical, the number of grid points has to be small enough that the runs can be performed quickly. However, a minimum grid size is needed to guarantee simulations of acceptable quality. The short runs are based on a corotating (i.e., tidally-locked) BNS system. For the long runs, we use the BNS labeled case $B$ in table II of \cite{Marronetti:2003hx}, which corresponds to an irrotational system outside the ISCO. The details for both grids are presented in table \ref{table_grids}. The number of points in the short run grid is about 73 times smaller than that of the long run simulations, making it possible to simulate a BNS orbital period in a couple of hours on a typical single-processor workstation. The long runs were performed using the IBM p690 Regatta cluster at NCSA.

\begin{table}
\caption{\label{table_grids} BNS Testing: Physical and numerical characteristics of the ID sets and Cartesian grids used for testing. The spatial resolution is given in number of grid points across the stellar diameter. The bounding box length $B$ gives the extent of the physical space covered in each direction (i.e.; the numerical grid spans from$[~-B,~0.0,~0.0]$ to $[~B,~B,~B]$ since we make use of the equatorial and $\pi$ symmetries of the systems), in units of total rest mass $M_{b0}$. The stars where modeled using a polytropic EOS with index n=1 ($\Gamma = 2$). For this particular EOS, the critical rest mass of a star in isolation is $m_{b0} = 0.180$. The binaries are composed of identical stars with individual rest masses with 80\% of the critical value. The last row gives the number of side-to-side light crossing times (lct) per orbital period.}
\begin{indented}
\item[]\begin{tabular}{@{}lcc}
\br
~ &  Short Runs~~~ & Long Runs\\
\mr
Binary type 			& corotating & irrotational   \\
Total Rest Mass $M_{b0}$        	& 0.2920 & 0.2932 \\
Total Grav. Mass $M_0$        	& 0.2697 & 0.2706 \\
Orbital ang. vel. $\Omega ~M_{b0}$	& 0.03456 & 0.02636 \\
Total Ang. Mom. $J_0 / M_0^2$		& 1.1469 & 1.0117 \\
Spatial Resolution       	& 20     & 40    \\
Grid Bounding Box $B / M_{b0}$ 	& 8.7    & 18.6  \\
Grid points              & $31^2 \times 60$ & $128^2 \times 256$ \\
lct per Orbital Period   	& 10.6   & 6.5   \\
\br
\end{tabular}
\end{indented}
\end{table}


\section{Results}
\label{results}

\subsection{Short Trial-and-Error Runs}
\label{QRC}

In this section we compare BSSN and Hamiltonian relaxation results using short trial-and-error runs (see table \ref{table_grids}). We concentrate here on the study of the Hamiltonian relaxation technique. A more comprehensive study of gauge fields choices and boundaries conditions will be done in the future. The BSSN benchmark run uses the K-Driver condition for the lapse and the $\Gamma$-Driver for the shift. The Hamiltonian relaxation method was tested using the same choice of gauge fields, plus the $\beta$-freeze condition. Apart from the use of the Hamiltonian relaxation for determining the conformal factor (with its respective boundary condition), the most important difference between this method and BSSN is the use of frozen boundary conditions for the lapse as explained in section \ref{BCs}. The results obtained with Hamiltonian relaxation in combination with the $\beta$-freeze condition are marginally better than those obtained with the $\Gamma$-Driver algorithm. Previous BNS simulations \cite{Duez:2002bn,Marronetti:2003hx} show that the $\Gamma$-Driver shift condition outperforms the $\beta$-freeze condition. In this article, we compare the combination BSSN/$\Gamma$-Driver vs. Hamiltonian Relaxation/$\beta$-freeze, given that those combinations are the best performers of each formulation.

Figure \ref{HC_BSSN_Evol} presents the evolution of the Hamiltonian residual $\mathcal{H}$ for the first three time steps of a BSSN simulation. The same curves are plotted on two different scales, to appreciate the effects on the bulk and the boundaries of the grid. We note that, while the residual remains unchanged in the bulk of the grid, the violation noise caused by Sommerfeld-like boundary conditions grows very fast at the boundaries. This effect is accentuated in the short runs, due to the small grid size. Figure \ref{HC_HR_Evol} shows the results of the Hamiltonian relaxation run. The special boundary conditions of the conformal factor $\psi$ control very effectively the violation noise generated at the grid's edge. At the same time, the Hamiltonian relaxation method reduces the constraint violation in the bulk of the grid.

The effect of the Hamiltonian relaxation technique on the overall quality of the run can be better appreciated by looking at the evolution of the total angular momentum of the BNS shown in figure \ref{J_BSSN_vs_HR_Short_Runs}. The total angular momentum is the most critical quality control curve in binary simulations. Any degradation in the quality of the run is appreciated first in the behavior of $J$. This is the case even for Newtonian simulations of binary systems \cite{Swesty:1999ke}. The angular momentum of the BSSN run is displayed for comparison (dashed line). The BSSN run performs reasonably well for about 1/3 of an orbital period before it becomes unstable. The Hamiltonian relaxation simulation continues, with degrading quality, for over an orbital period. This is a remarkable result considering the small size of the short run grid, where an orbital period is roughly equivalent to 10 side-to-side light crossing times.

\begin{figure}
\epsfxsize=4.0in
\begin{center}
\leavevmode \epsffile{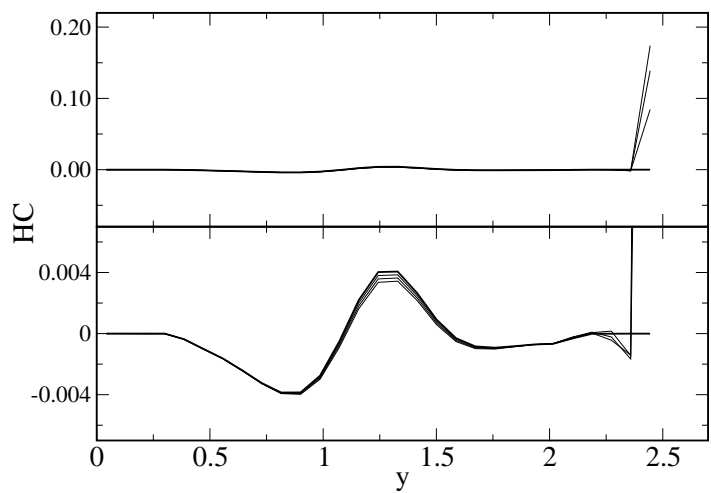}
\end{center}
\caption{ \underline{BSSN Short Run}: Evolution of the Hamiltonian constraint residual for the first three time steps. The same curves are plotted on two different scales, to appreciate the effects on the boundaries (top) and the bulk (bottom) of the grid. The curves are plotted following the line with coordinates $(0,y,0)$, that runs through the center of the star. The companion star, located on the negative $y$ hemisphere, is not shown.}
\label{HC_BSSN_Evol}
\end{figure}

\begin{figure}
\epsfxsize=4.0in
\begin{center}
\leavevmode \epsffile{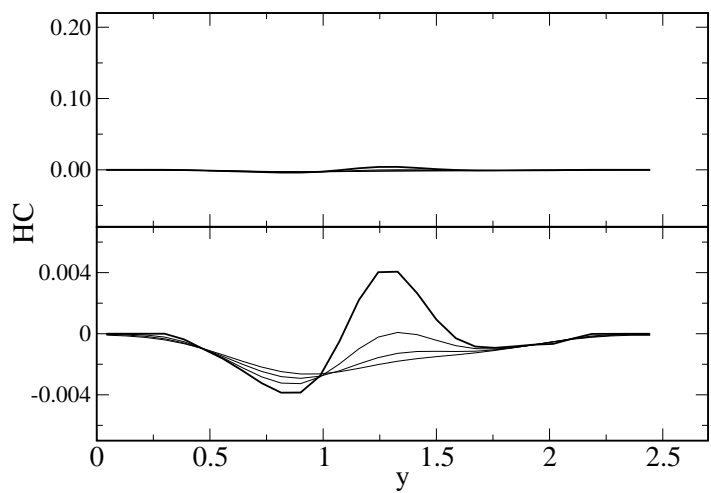}
\end{center}
\caption{ \underline{Hamiltonian Relaxation Short Run}: Evolution of the Hamiltonian constraint residual for the first three time steps.}
\label{HC_HR_Evol}
\end{figure}

\begin{figure}
\epsfxsize=4.0in
\begin{center}
\leavevmode \epsffile{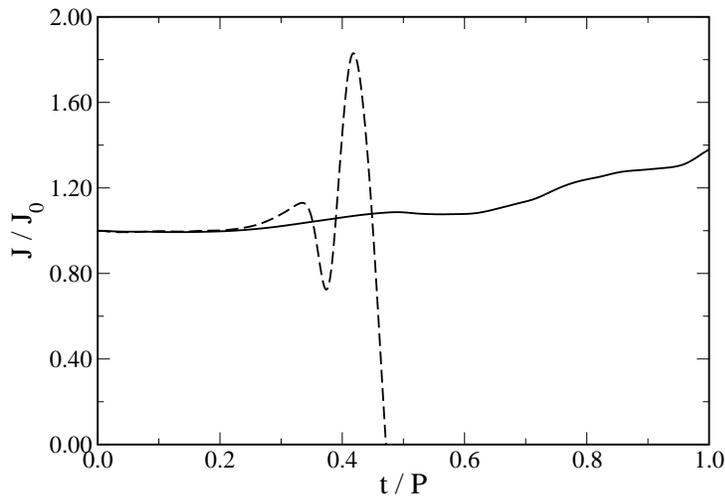}
\end{center}
\caption{ \underline{Short Runs}: Total angular momentum $J$ as a function of time, given as fraction of the orbital period $P$. $J$ is normalized to its initial value $J_0$. The solid (dashed) line corresponds to the Hamiltonian relaxation (BSSN) run.}
\label{J_BSSN_vs_HR_Short_Runs}
\end{figure}


\subsection{Long High-Resolution Large-Grid Runs}
\label{LRC}

The Hamiltonian relaxation method was tested with a more accurate simulation, by increasing the size and spatial resolution of the numerical grid and starting from an ID set corresponding to an irrotational binary (see table \ref{table_grids}). We discuss in this section the most important features of such simulation, while the corresponding convergence tests are covered in \ref{appendix_CT}. All the plots of this section show curves that, for clarity, have been normalized to their corresponding initial values. The solid (dashed) lines represent the results for the Hamiltonian relaxation (BSSN) run.

The evolution of the $L_2$ norm of the Hamiltonian constraint residual $\mathcal{H}$ across the numerical grid is shown in figure \ref{HC_BSSN_vs_HR}. The satisfaction of the constraint improves by about two orders of magnitude with respect to the BSSN run. The suppression of the constraint violation is such that, in average, the Hamiltonian residual is five times smaller than the ID set value. This drastic reduction of the constraint violation suggests a new way of generating ID for binaries: a snapshot of all the gravitational and hydrodynamical fields taken at $t/P = 0.5$  can be used as an ID set for evolutions with different numerical codes. Note that the ID set obtained in this manner would not be subject to the conformal flatness restriction for the spatial metric. On the contrary, given that (for this grid) half a period is equivalent to more than 3 side-to-side light crossing times, the ID set would possess the characteristics (to some extent) of outgoing gravitational radiative fields.

Figure \ref{J_BSSN_vs_HR} shows the evolution of the total angular momentum. In addition to the Hamiltonian relaxation and BSSN curves, we plotted the PN estimation (dotted line) of the angular momentum loss for a point-mass binary with the same mass and angular momentum as the BNS in consideration (see \ref{appendix_PN}). The Hamiltonian relaxation run agrees with the PN prediction for about 1.5 orbital periods, while the BSSN curve starts going upward well before the first period. The inset of figure \ref{J_BSSN_vs_HR} zooms in on the first half a period, showing the reduced level of noise in the Hamiltonian relaxation curve.

Figure \ref{QC_BSSN_vs_HR} shows the remaining quality control curves: the coordinate separation between stellar centers $d$, the total gravitational $M$ mass, and the $x$ component of the momentum constraint. The total rest mass of the system remains invariant to within a 0.1 \% in both runs. 

\begin{figure}
\epsfxsize=4.0in
\begin{center}
\leavevmode \epsffile{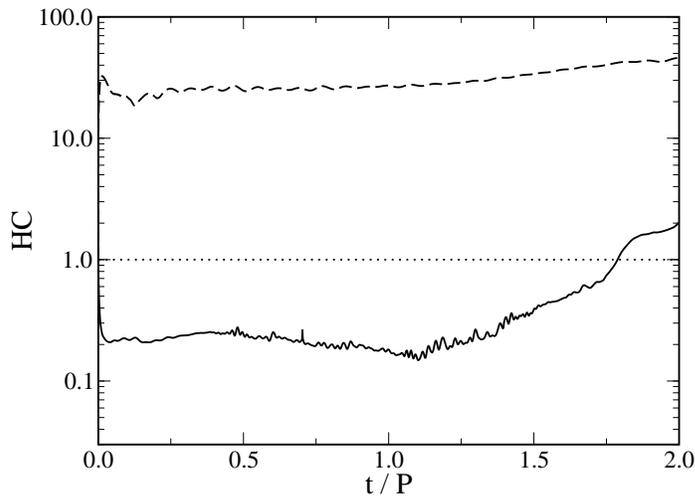}
\end{center}
\caption{ \underline{Long Runs}: Evolution of the $L_2$ norm of the Hamiltonian constraint violation across the numerical grid. The solid (dashed) line corresponds to the Hamiltonian relaxation (BSSN) run. The dotted line marks the violation at the initial time step. Note that the curves are plotted in logarithmic scale to highlight the more than two orders of magnitude difference between the Hamiltonian relaxation and BSSN results. The Hamiltonian relaxation scheme not only suppresses the constraint violation modes, but also reduces the violation present in the ID set by a factor of about 5.}
\label{HC_BSSN_vs_HR}
\end{figure}

\begin{figure}
\epsfxsize=4.0in
\begin{center}
\leavevmode \epsffile{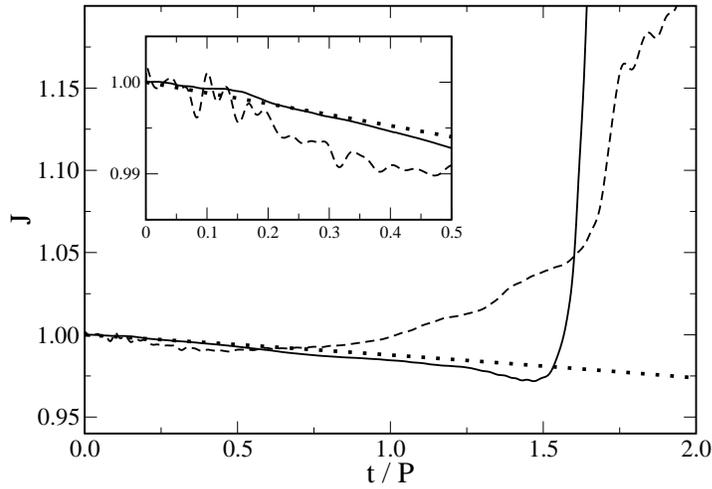}
\end{center}
\caption{ \underline{Long Runs}: Evolution of the total angular momentum $J$. The solid (dashed) line corresponds to the Hamiltonian relaxation (BSSN) run, while the dotted line shows the PN estimation (see \ref{appendix_PN}). The inset expands the plot for the first half orbital period.}
\label{J_BSSN_vs_HR}
\end{figure}

\begin{figure}
\epsfxsize=4.0in
\begin{center}
\leavevmode \epsffile{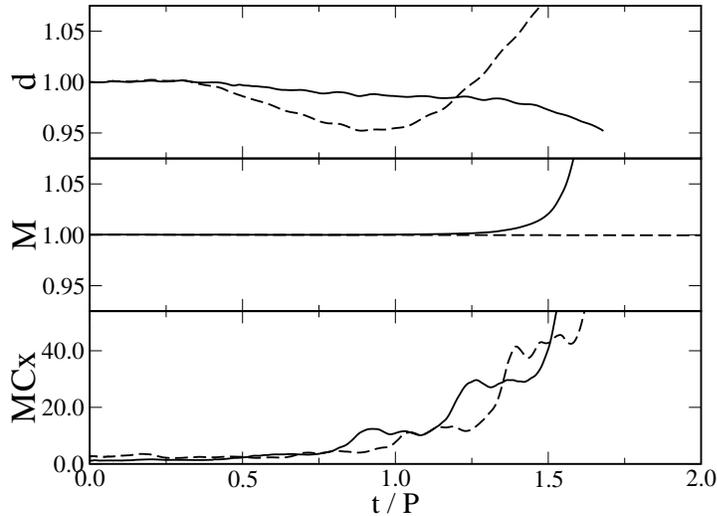}
\end{center}
\caption{ \underline{Long Runs}: Remaining quality control curves for Hamiltonian relaxation (solid lines) and BSSN (dashed lines). From top to bottom, we show the evolution of the coordinate separation between stellar centers $d$, the total gravitational $M$ mass, and the $x$ component of the momentum constraint.}
\label{QC_BSSN_vs_HR}
\end{figure}

We note that the Hamiltonian relaxation run degrades quickly after 3/2 orbital periods (about 10 side-to-side light crossing times), while the BSSN run continues for over another period before stopping. However, the overall quality of the Hamiltonian relaxation results is superior to those of the BSSN simulation during the overlapping time. The onset of the instability at the end of the simulation seems to be independent on the grid size (see figure \ref{J_HR_Conv} in \ref{appendix_CT}), possibly indicating that the problem is not originated at the boundaries. The total mass and angular momentum of the binary are affected much earlier than the Hamiltonian constraint violation, which may indicate that the relaxation technique is not at fault either. One possible cause for the premature end of the simulation could be our choice of $\beta$ freezing for the shift vector, which is known to perform poorly when the matter distribution drifts significantly from the initial configuration. This instability will be more rigorously studied in future work.


\section{Conclusions}
\label{conclusions}

We introduced a new testing ground for numerical relativistic formulations and finite-difference techniques, based on short trial-and-error runs of BNS systems. We show the usefulness of BNS testing by developing a new technique based of the approximate solution of the Hamiltonian constraint (Hamiltonian relaxation). The new method was tested using more realistic and computationally demanding numerical simulations of BNS outside the ISCO, and the results were compared with those from BSSN simulations. The Hamiltonian relaxation improves the overall quality of the simulations by reducing the Hamiltonian constraint violation by about two orders of magnitude with respect to BSSN values, and by a factor of five with respect to the violation present in the initial data set. More remarkably, the improvement in the time evolution of the total angular momentum of the binary is such that it agrees with PN point-mass binary estimations. The Hamiltonian relaxation run becomes unstable at about 3/2 orbital periods ($\sim$ 10 side-to-side light crossing times), possibly due to the use of $\beta$ freezing shift condition. This problem will be addressed in future work. Similarly, we will attempt to generalize the Hamiltonian relaxation technique to the momentum constraint, to further improve the quality of the simulations. Finally, note that the Hamiltonian relaxation method can easily be applied to other formulations other than BSSN, since only requires the conformal decomposition of the spatial metric tensor. This subject will also be explored in the future.


\ack
It is a pleasure to thank P. Laguna, G. Mathews, and W. Miller for helpful discussions. Special thanks to C. Finkel for carefully reading the manuscript. This work was partially supported by National Computational Science Alliance under Grants PHY020007N. PHY050010T, and PHY050015N. It is also a pleasure to acknowledge the help provided by the Cactus Code organization with the setup and running of the Cactus environment on different platforms.


\begin{appendix}

\section{Code Tests}
\label{appendix_CT}

\subsection{Validation Tests}

\begin{figure}
\epsfxsize=4.0in
\begin{center}
\leavevmode \epsffile{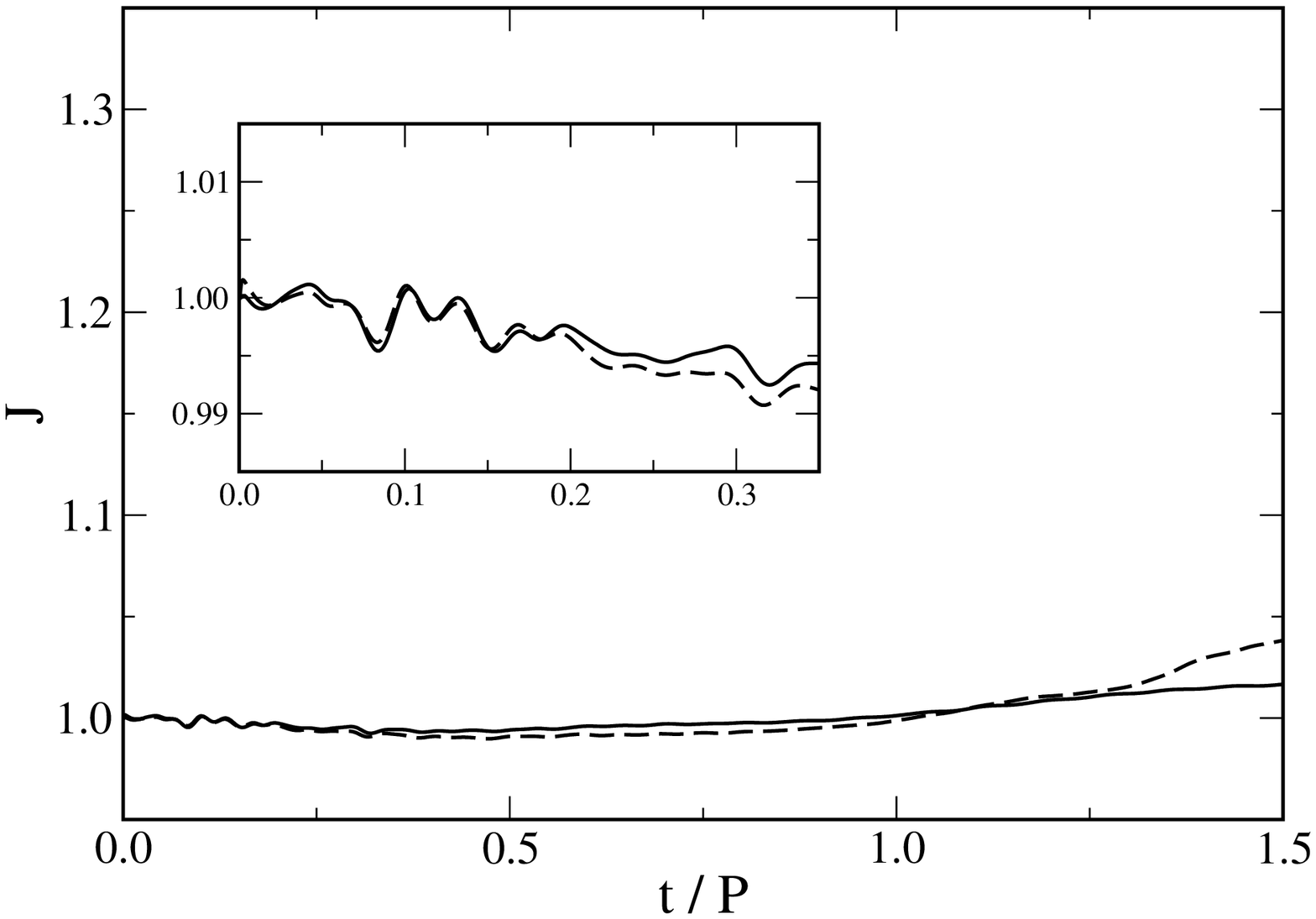}
\end{center}
\caption{ Angular Momentum comparison between GRHyd (dashed line) and results from Marronetti \et \cite{Marronetti:2003hx} (solid line). The GRHyd simulation starts with the long run ID set of table \ref{table_grids} which corresponds to case $B$ of table II in \cite{Marronetti:2003hx}. The inset zooms in on the first part of the evolution.}
\label{J_ABE_vs_GRHyd}
\end{figure}

Several code validation runs were performed with the new general relativistic code GRHyd. Using GRHyd's BSSN and van Leer modules (see section \ref{GRHyd}) and employing the same ID sets, we performed BNS simulations that were compared to those in \cite{Marronetti:2003hx}.

Figure \ref{J_ABE_vs_GRHyd} compares the evolution of the total angular momentum of the long run simulation described in table \ref{table_grids}, with the corresponding case $B$ of table II in \cite{Marronetti:2003hx}. The solid (dashed) line corresponds to the results from \cite{Marronetti:2003hx} (GRHyd). The inset zooms in on the first part of the simulation, to highlight the agreement between both codes. The differences between the runs are due to two sources: roundoff error, which can cause drifts of the order of a fraction of a percent in extensive quantities like the total angular momentum, and, more importantly, a grid size difference between both runs. The numerical grid employed by GRHyd is one grid zone shorter than the one used in \cite{Marronetti:2003hx}. This is due to the different ways in which both codes setup the symmetries in the numerical grid. This effect can be corrected by generating a larger-grid ID set and using it for new simulations with both codes. For clarity, we decided to maintain the same ID set used in \cite{Marronetti:2003hx}. The results start diverging significantly from each other after 1.3 orbital periods. This is expected, since the boundary effects are by then dominant, degrading significantly the quality of the simulation.

\subsection{Convergence Tests}

We tested the convergence of the results obtained with the Hamiltonian relaxation method with the size of the numerical grid. We performed the long run simulation of table \ref{table_grids} on four different grids with lengths $B/M_{b0}=9.3$,  $11.6$, $14.0$, and $18.6$. All grids have the same spatial resolution of about 40 points across the stellar radius. The results are presented in figure \ref{J_HR_Conv}, where we show the total angular momentum as a function of time. The inset expands the scale for the first half orbital period. The plot shows the convergence of the numerical results towards the PN estimation for point-mass binaries. The agreement between the three largest grids' results indicates that they could be used for preliminary runs, for studies which require extensive parameter space searches, like in the determination of the BNS ISCO (see \cite{Marronetti:2003hx}). 

We also tested the numerical convergence of the Hamiltonian relaxation results with the spatial grid resolution. We performed three runs based on the irrotational long run of table \ref{table_grids} using 20 (low res.), 25 (medium res.), and 40 (high res.) grid points across the stellar diameter. Figure \ref{J_HR_Conv2} shows the normalized total angular momentum, where the separation between curves is consistent with second order convergence in grid spacing. 

In order to see the second order convergence of the Hamiltonian constraint violation with grid resolution the relaxation stopping criterion needs to be changed: instead of comparing the $L_2$ of the Hamiltonian constraint residual with its value at the previous time step (see section \ref{TEE}), we compare it to the value at the previous {\it iteration} step (represented by the index $m$ in equation \ref{update}). The relaxation is stopped when the difference between these norms falls below some threshold value. The results for the three different grid resolutions are shown in figure \ref{H_HR_Conv2}. The change in criterion has little effect on the results (mass and angular momentum do not vary significantly), however the computational time is considerably longer.

\begin{figure}
\epsfxsize=4.0in
\begin{center}
\leavevmode \epsffile{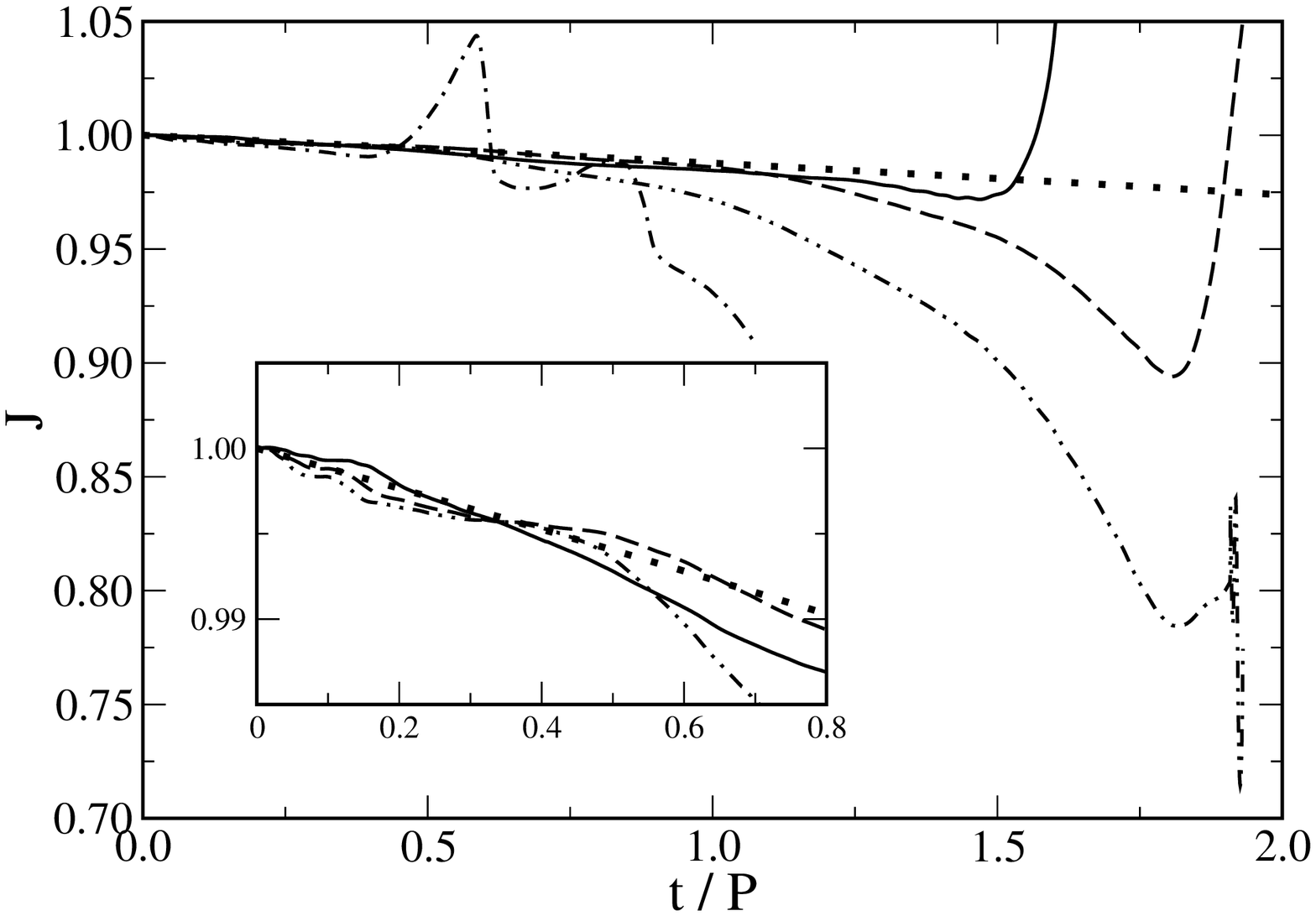}
\end{center}
\caption{ Convergence of the Hamiltonian relaxation results with varying grid sizes. The convergence test is based on the long run simulation (see table \ref{table_grids}) and the plot shows the behavior of the total angular momentum. The labels for curves are (from smallest to largest): dash-dotted, dash-double-dotted, dashed, and solid. All the grids have the same spatial resolution. The dotted line shows the PN estimation (see \ref{appendix_PN}). The inset expands the plot for the first half of the run.}
\label{J_HR_Conv}
\end{figure}

\begin{figure}
\epsfxsize=4.0in
\begin{center}
\leavevmode \epsffile{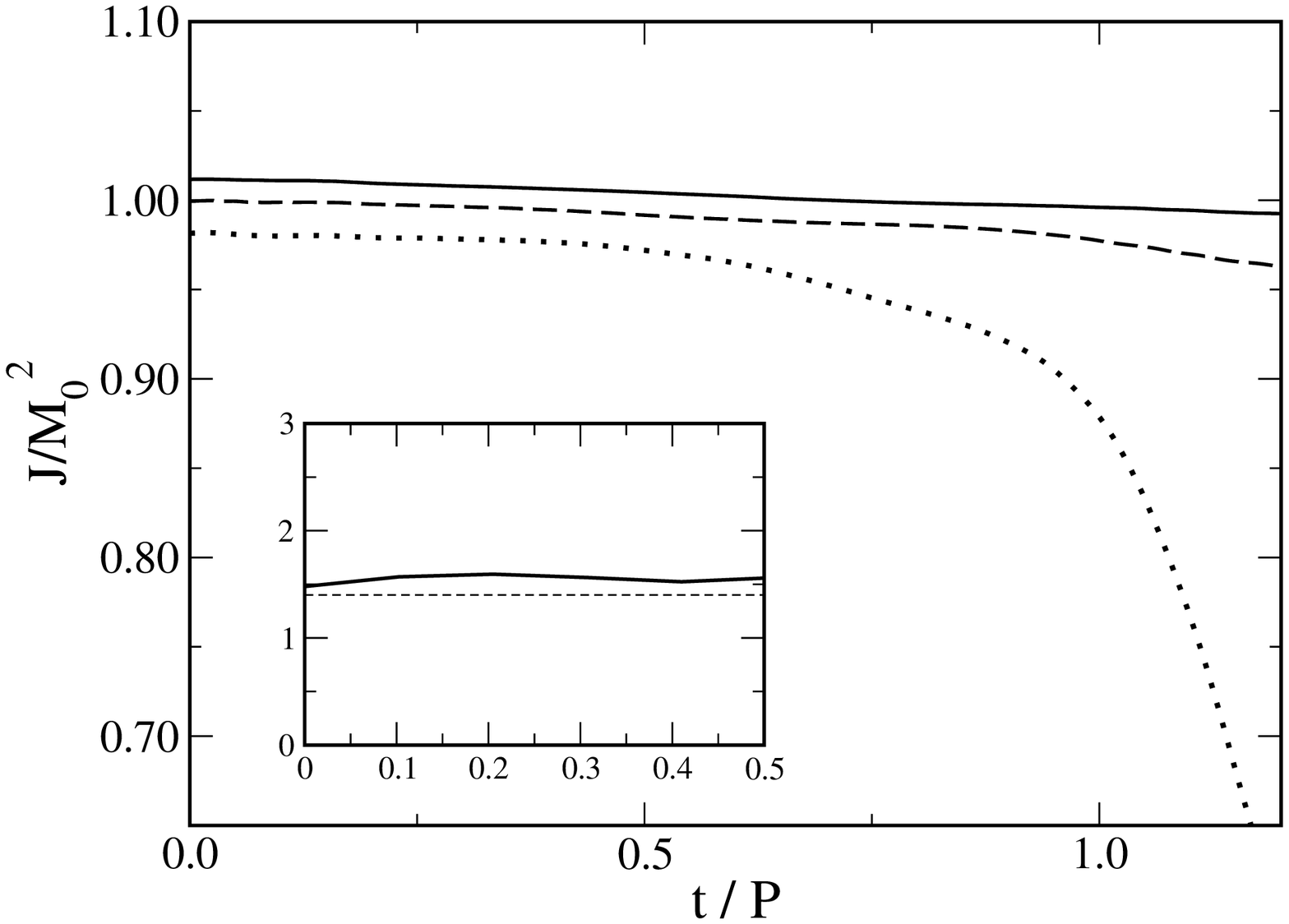}
\end{center}
\caption{ Total angular momentum in units of total ADM mass squared for the irrotational Long Run described in table \ref{table_grids} with three different grid resolutions: low (dotted line), medium (dashed line), and high (solid line). The inset shows the ratio between the curves $(Med-Low)/(High-Med)$ which, for second order convergence, is expected to be about 1.40 (dashed line).}
\label{J_HR_Conv2}
\end{figure}

\begin{figure}
\epsfxsize=4.0in
\begin{center}
\leavevmode \epsffile{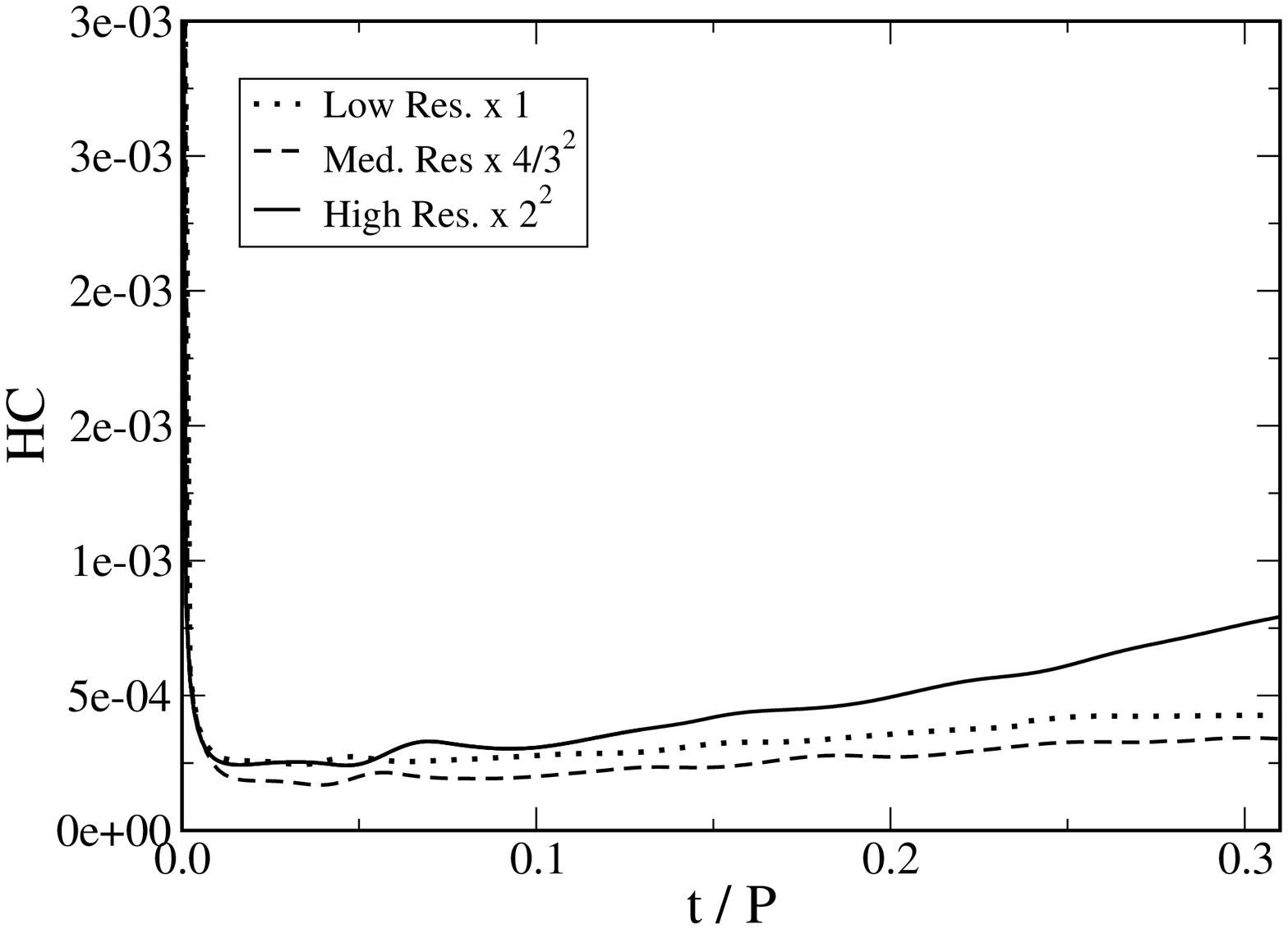}
\end{center}
\caption{ Evolution of the $L_2$ norm of the violation of the Hamiltonian constraint for the irrotational Long Run described in table \ref{table_grids} with three different grid resolutions. Curve labeling is the same as in figure \ref{J_HR_Conv2}. The numerical factors multiplying the curves correspond to the ratios between grid spacings.}
\label{H_HR_Conv2}
\end{figure}


\section{Post-Newtonian Estimation of Angular Momentum Loss}
\label{appendix_PN}

In order to get an independent estimation of the angular momentum loss due to the gravitational radiation, we calculated the post-Newtonian prediction corresponding to a point-mass binary in circular orbit. Kidder \cite{Kidder:1995} derives the 2.5 PN equations of motion for such binaries with arbitrary masses and spins in quasi-circular orbits, which is a good approximation when the inspiral motion following the radiation timescale is much larger than the orbital motion. For equal-mass zero-spin stars the loss of angular momentum becomes
\ba
\frac{\rmd J}{\rmd t} = -\frac{32}{5} \eta^2 ~ 	\left(\frac{M_0}{r}\right)^4 (M_0
r)^{1/2}  \left[ 1 - \frac{2423+588 \eta}{336} \left(\frac{M_0}{r}\right) + 4 \pi \left(\frac{M_0}{r}\right)^{3/2} \right], \nonumber
\ea
where $M_0$ is the total mass, $\eta \equiv M_1 M_2 /M_0^2$ the reduced mass, $r$ the coordinate separation, and $J$ the total angular momentum. This formula was derived following the Symmetric Trace-Free (STF) radiative multipoles treatment given by Thorne \cite{Thorne:1980}. It requires a formula for the point-mass coordinate separation $r$ as a function of time, which can be calculated from the equations of motion as 
\ba
\frac{\rmd r}{\rmd t}  = -\frac{64}{5} ~ \eta ~ \left(\frac{M_0}{r}\right)^3 
\left[ 1 - \frac{1751+588 \eta}{336} \left(\frac{M_0}{r}\right)+4 \pi ~ \left(\frac{M_0}{r}\right)^{3/2} \right]. \nonumber
\ea
The dotted curves presented in Figs. \ref{J_BSSN_vs_HR} and \ref{J_HR_Conv} were obtained by integrating these ODEs, assuming that the inspiral starts from a stationary circular orbit, with initial values for $M_0$, $J_0$, and the angular velocity $\Omega$ given in table \ref{table_grids}. The initial coordinate separation $r_0$ is determined by Kepler's law
\ba
M_0 \Omega = \left(\frac{M_0}{r_0}\right)^{3/2}~. \nonumber
\ea
\\

\end{appendix}


\end{document}